# Reciprocity breaking during nonlinear propagation of adapted beams through random media

J.P. Palastro[1], J. Peñano[1], W. Nelson[2], G. DiComo[2,3], L. A. Johnson[1], M.H. Helle[1], and B. Hafizi[1]

[1]*Naval Research Laboratory, Washington DC 20375-5346, USA*
[2]*Department of Electrical Engineering, University of Maryland, College Park MD 20740, USA*
[3]*Research Support Instruments, Lanham MD 20706, USA*

**Abstract**

Adaptive optics (AO) systems rely on the principle of reciprocity, or symmetry with respect to the interchange of point sources and receivers. These systems use the light received from a low power emitter on or near a target to compensate profile aberrations acquired by a laser beam during linear propagation through random media. If, however, the laser beam propagates nonlinearly, reciprocity is broken, potentially undermining AO correction. Here we examine the consequences of this breakdown. While discussed for general random and nonlinear media, we consider specific examples of Kerr-nonlinear, turbulent atmosphere.

Optical configurations often exhibit reciprocity, or symmetry with respect to the interchange of point sources and receivers [1-3]. It is precisely this symmetry that enables adaptive optics (AO) correction of laser beam profiles delivered to targets in random media. AO correction uses the light received from a low power emitter, or beacon, on or near the target to adjust the laser beam's spatial profile [2,4-9]. In a reciprocal configuration, every 'ray' in the beacon has a reciprocal partner in the beam. These rays traverse the random media along identical paths but in opposite directions. Thus by reversing the rays along the phase front, or phase conjugating, the beacon irradiance profile can be reproduced at its source. Often, however, the rays on the incoming and outgoing paths experience differing dielectric environments. The medium evolves, or as is the interest here, the power in the beam surpasses that of the beacon, leading to differences in the nonlinear refraction on the outgoing and incoming paths.

Here we examine the nonlinear breakdown of reciprocity occurring when a low power beacon informs the phase correction of a high peak power laser beam. We introduce a metric, an overlap of the beacon and the beam fields, that quantifies the breakdown, and provides a necessary and sufficient condition for reciprocity. The metric is applied to the specific case of field conjugated high power beams propagating through Kerr-nonlinear turbulent atmosphere. The degree of overlap, henceforth referred to as reciprocity, is found to drop rapidly at powers approaching the critical power for self-focusing. In the strong turbulence, the reciprocity increases due to spatial incoherence weakening self-focusing. A rough scaling, explaining this behavior, is derived. Finally, we find that the drop in reciprocity is dominated by phase differences between the beacon and beam, suggesting that AO correction can be effective when the on-target irradiance is important, but not the phase.

While there are several types of beacons and variations on AO implementations [2,4-9], we consider a simple optical configuration that illustrates the salient physical phenomena. The configuration is displayed in Fig. 1. A static random medium separates the target plane on the right from the receiver plane on the left. The beacon resides in the target plane, and the receiver plane coincides with the laser beam transmitter plane. The beacon light propagates through the random medium and is collected in the receiver plane where its phase and amplitude are measured. The conjugate phase and amplitude are then applied to a laser beam, which propagates back to the target through the same random medium. In Fig. 1 the different colors of the beacon and laser beam are for illustrative purposes only; their wavelengths, in actuality, would be quite similar.

To model the light propagation, we use the scalar paraxial wave equation. We note, however, that the conceptual discussion of reciprocity and its breakdown applies to other wave equations as well, including the vector and scalar Helmholtz equations. The transverse electric field, $E_\perp$, consists of a carrier wave modulated by a slowly varying envelope, $E$: $E_\perp(\mathbf{x},t) = \tfrac{1}{2} E(\mathbf{x}) \exp[i(kz - \omega t)] + \text{c.c.}$ where $\omega$ is the carrier frequency, $k = \omega n_0 / c$, and $n_0$ is a reference refractive index. The envelope evolves according to

$$\frac{\partial}{\partial z} E(\mathbf{x}) = i \frac{1}{2k} \left[ \nabla_\perp^2 + 2k^2 n_0 \delta n(\mathbf{x}) \right] E(\mathbf{x}) \qquad (1)$$

where $\delta n(\mathbf{x}) = n(\mathbf{x}) - n_0$ is the refractive index shift and $n(\mathbf{x})$ the total refractive index. The refractive index shift consists of spatially dependent linear and nonlinear components. The linear component, $\delta n_L = \delta n_i + \delta n_f$, accounts for gain or dissipation, $\delta n_i$, and random fluctuations in the medium, $\delta n_f$. The fluctuations have zero mean when averaged over an ensemble of statistically independent instances. The nonlinear component, $\delta n_{NL}(I)$, is a function of the intensity, $I = \tfrac{1}{2} c \varepsilon_0 n_0 |E(\mathbf{x})|^2$. Explicitly, $\delta n = \delta n_i + \delta n_f + \delta n_{NL}$ with $\text{Im}[\delta n] = \delta n_i$.

In the following, we make use of the Green's functions for Eq. (1). In particular, we define

$$\left[ i \frac{\partial}{\partial z} + \frac{1}{2k} \nabla_\perp^2 + k n_0 \delta n_L \right] G^+(\mathbf{r}, z; \mathbf{r}', z') = \delta(\mathbf{x} - \mathbf{x}') \qquad (2a)$$

$$\left[ -i \frac{\partial}{\partial z} + \frac{1}{2k} \nabla_\perp^2 + k n_0 \delta n_L \right] G^-(\mathbf{r}, z; \mathbf{r}'', z'') = \delta(\mathbf{x} - \mathbf{x}'') \qquad (2b)$$

where $G^+(\mathbf{r},z;\mathbf{r}',z')$ and $G^-(\mathbf{r},z;\mathbf{r}'',z'')$ propagate the field when $z>z'$ and $z<z''$ respectively. While we never calculate it explicitly, the Green's function provides a succinct description of reciprocity. Multiplying Eq. (2a) by $G^-(\mathbf{r},z;\mathbf{r}'',z'')$ and Eq. (2b) by $G^+(\mathbf{r},z;\mathbf{r}',z')$, subtracting the results, and integrating over all space, we find the reciprocity relationship $G^+(\mathbf{r},z;\mathbf{r}',z') = G^-(\mathbf{r}',z';\mathbf{r},z)$: the linear optical configuration is symmetric with respect to the interchange of point sources and receivers. This symmetry holds even in the presence of gain and dissipation. In the absence of these, $\delta n_i = 0$, one can follow a similar derivation to demonstrate the equivalence of reciprocity and reversibility in the axial coordinate: $G^{+*}(\mathbf{r},z;\mathbf{r}',z') = G^-(\mathbf{r},z;\mathbf{r}',z')$. Reversibility implies reciprocity, but the converse is not true.

We continue by describing an idealized AO system that illustrates the importance of reciprocity [2]. We denote the beacon and laser beam electric field envelopes as $E_B(\mathbf{r},z)$ and $E_L(\mathbf{r},z)$ respectively. The receiver/transmitter resides at $z=0$ and the target at $z=z_T$. For this example, we take $\delta n_i = 0$; when $\delta n_i = \delta n_i(z)$, the amplitudes can be adjusted retroactively by the appropriate exponential factor, $\exp[k\int \delta n_i(z)dz]$. The Green's function $G^-(\mathbf{r},0;\mathbf{r}',z_T)$ propagates the beacon field from the target to the receiver: $E_B(\mathbf{r},0) = i\int G^-(\mathbf{r},0;\mathbf{r}',z_T)E_B(\mathbf{r}',z_T)d\mathbf{r}'$. At the receiver, the AO system applies the conjugate profile to the outgoing beam, $E_L(\mathbf{r},0) = -i\int G^{-*}(\mathbf{r},0;\mathbf{r}',z_T)E_B^*(\mathbf{r}',0)d\mathbf{r}'$. The Green's function $G^+(\mathbf{r},z_T;\mathbf{r}',0)$ propagates the beam from the transmitter to the target: $E_T(\mathbf{r},z_T) = i\int G^+(\mathbf{r},z_T;\mathbf{r}',0)E_L(\mathbf{r}',0)d\mathbf{r}'$, which upon substitution of the outgoing beam field provides $E_T(\mathbf{r},z_T) = \iint G^+(\mathbf{r},z_T;\mathbf{r}',0)G^{-*}(\mathbf{r}',0;\mathbf{r}'',z_T)E_B^*(\mathbf{r}'',z_T)d\mathbf{r}''d\mathbf{r}'$. If the channel is reciprocal, the on-target beam field reduces to $E_L(\mathbf{r},z_T) = E_B^*(\mathbf{r},z_T)$. The AO system has exploited reciprocity to illuminate the target with the conjugate field of the beacon.

The nonlinear refractive index, $\delta n_{NL}$, was excluded in this example, and neither the beacon nor the beam propagated nonlinearly. Moreover, the Green's function, a linear construct, was used to define the conditions of reciprocity and reversibility. To demonstrate nonlinear reciprocity and reversibility, we divide propagation over a total distance $L$ into $N$ steps of size $\Delta z = L/N$. Forward and backward propagation over a single step are expressed as

$$E_z(\mathbf{r}) = \int H^{\mp}(\mathbf{r},z;\mathbf{r}'',z\pm\Delta z) E_{z\pm\Delta z}(\mathbf{r}'') d\mathbf{r}'', \qquad (3a)$$

$$H^{\mp}(\mathbf{r},z;\mathbf{r}'',z\pm\Delta z) = -\int G^{\mp}(\mathbf{r},z;\mathbf{r}',z\pm\tfrac{\Delta z}{2}) e^{ik\Delta z \delta n_{NL}^h} G^{\mp}(\mathbf{r}',z\pm\tfrac{\Delta z}{2};\mathbf{r}'',z\pm\Delta z) d\mathbf{r}' \qquad (3b)$$

where $\delta n_{NL}^h = \delta n_{NL}[I(\mathbf{r}',z\pm\tfrac{\Delta z}{2})]$, and $G^+$ and $G^-$ are defined as before. Successive application of the integral in Eq. (3a) propagates the envelope over multiple steps. It is clear from Eq. (3b) that if the linear configuration is reciprocal, then $H^+(\mathbf{r},z;\mathbf{r}',z-\Delta z) = H^-(\mathbf{r}',z-\Delta z;\mathbf{r},z)$, and if it is reversible then $H^{+*}(\mathbf{r},z;\mathbf{r}',z-\Delta z) = H^-(\mathbf{r},z;\mathbf{r}',z-\Delta z)$. By using these relations for $H^{\pm}$ in an expression where Eq. (3a) is successively applied and taking the limit of infinitesimal $\Delta z$, one can show that a nonlinear configuration with real intensity dependent refractive is reciprocal or reversible.

This nonlinear reciprocity can be applied to our AO example when the beacon and beam experience identical optical configurations. From a practical standpoint, however, the propagation of the beacon light from the target to the receiver, and the propagation of the beam from the transmitter to the target can occur under different conditions. The random media may change in time or, as is the interest here, the power of the beacon and beam may differ. This results in an effective breakdown of reciprocity. The symmetry breaking can be expressed symbolically by parameterizing $H^{\pm}$ with the beacon and beam powers, $H^+(\mathbf{r},z;\mathbf{r}',z-\Delta z;P_L) \neq H^-(\mathbf{r}',z-\Delta z;\mathbf{r},z;P_B)$ where $P_j = \int I_j d\mathbf{r}$. Conceptually, nonlinear refraction causes the beacon and beam rays to take different paths through the medium.

In order to quantify the breakdown of reciprocity along the propagation path, we define the following metric:

$$R(z) \equiv \frac{1}{2} \frac{\varepsilon_0 c}{[P_B(z) P_L(z)]^{1/2}} \int E_B(\mathbf{r},z) E_L(\mathbf{r},z) d\mathbf{r}, \qquad (4)$$

where $|R|\leq 1$. Equation (4) is simply the overlap of the beam and beacon fields. The normalization was chosen such that if the beam field is everywhere the conjugate of the beacon field, $R=1$. As a result, the criterion $R(z)=1$ for all $z$ provides a necessary and sufficient condition for reciprocity of an optical configuration.

A few examples aid in the interpretation of $R$. First consider the idealized AO system discussed above. When the beacon and beam have identical powers, we showed that $E_L(\mathbf{r},z_T)=E_B^*(\mathbf{r},z_T)$, which can be straightforwardly generalized to $E_L(\mathbf{r},z)=E_B^*(\mathbf{r},z)$. Inserting this field into Eq. (4), we find $R(z)=1$ for all $z$. Suppose $E_L$ and $E_B$ have identical amplitudes but are everywhere phase shifted by $\pi/2$ ($\pi$), then $R=i$ ($-1$): $\arg(R)\neq 0$ always indicates a phase difference. If $E_L$ and $E_B$ are spatially disjoint, implying that their 'rays' propagate through wholly different regions of the random medium, then $R=0$. A value of $|R|<1$ does not, however, indicate a unique spatial phase difference and irradiance disjointedness.

To demonstrate application of this metric, we simulated the optical configuration illustrated in Fig. 1 for the case in which the random media is dissipationless, Kerr-nonlinear, turbulent atmosphere. The Kerr nonlinearity, $\delta n_{NL}(I)=n_2 I$ where $n_2$ is second order nonlinear refractive index, permits the well-known phenomenon of self-focusing and beam collapse [10,11]. The ratio of the total beam power to the critical power, $P_{cr}\sim \lambda^2/2\pi n_0 n_2$, parameterizes the effect. For an initially collimated Gaussian beam with spot size $w$, the collapse distance in uniform media was developed by Marburger: $z_c = 0.18 kw^2/\{[(P/P_{cr})^{1/2}-0.85]^2-.022\}^{1/2}$ for $P>P_{cr}$ [10]. Here we limit the propagation to distances well less than $z_c$.

The simulation involves three steps. In the first step, the beacon field is propagated from the target to the receiver using Eq. (1) with $\delta n_f$ included as phase screens [11-13]. The modified Von Karman spectrum was used for the Fourier transform of $\delta n_f$'s covariance [13-15]. The second simulation step initiates the laser beam envelope with the conjugated and amplified receiver plane beacon envelope: $E_L(\mathbf{r},0)=\eta E_B^*(\mathbf{r},0)$ with

$\eta > 1$. In the third step, the beam is propagated to the target, encountering the same phase screens as the beacon at the appropriate axial positions.

In all of the simulations presented, the initial beacon field had a Gaussian profile, $E_B(\mathbf{r},0) = E_0 \exp(-r^2/w^2)$. The amplitude, $E_0$, was chosen such that the power, $P_B = \frac{\pi}{4} c\varepsilon_0 n_0 w^2 E_0^2$ was far below $P_{cr}$, ensuring linear propagation. The initial beam power, $P_L$, was varied from below $P_{cr}$ to above $P_{cr}$. Statistical quantities, such as ensemble averages, denoted by $\langle \rangle$, and standard deviations, were obtained by simulating the propagation through $10^3$ statistically independent realizations of the turbulence.

We considered an atmospheric propagation regime where four parameters are required for characterization: $P_L/P_{cr}$ which has already been discussed, the Rayleigh length $Z_R = \frac{1}{2} kw^2$, the Rytov variance $\sigma_r^2 = 1.2 C_n^2 k^{7/6} z^{11/6}$ where $C_n^2$ is the refractive index structure constant, and the turbulence inner scale length $\ell_0$. For simplicity, the propagation distance was limited to $z_T = 0.12 Z_R$, such that in the absence of index fluctuations the beacon would be collimated. The Rytov variance describes the normalized intensity variance of a plane wave, and provides a convenient metric for the optical turbulence strength [14,15]. In particular $\sigma_r^2 > 1$ provides a rough condition for strong optical turbulence. The ratio of the inner scale to the laser spot size determines the relative importance of beam spreading and wander, with wander dominating when $\ell_0/w \gg 1$. In these simulations, the inner scale length was fixed at $\ell_0 = w/8$.

Figure 2(a) displays the ensemble averaged $R(z_T)$ as a function of $P_L/P_{cr}$ for a turbulence strength of $\sigma_r^2 = 6.8$. The dots, squares, and triangles represent the means of $|R(z_T)|$, $\text{Re}[R(z_T)]$, and $\text{Im}[R(z_T)]$ respectively. The swath boundaries illustrate +/- the standard deviation of $|R(z_T)|$. The real (imaginary) component of $R(z_T)$ decreases (increases) with increasing beam power consistent with modified propagation of the beam due to nonlinear focusing. We return to the apparent scalings, $\text{Re}\langle R(z_T) - 1\rangle \propto -(P_L/P_{cr})^2$ and $\text{Im}\langle R(z_T)\rangle \propto (P_L/P_{cr})$ for $P_L/P_{cr} < 1.0$, below. The standard deviation of $|R(z_T)|$ increases with the beam power, demonstrating that, even when phase-corrected, high power beam propagation is sensitive to the specific realization of turbulence. As an example, Figs. 2(c) and (d) show two instances of on-target intensity profiles for a beam with $P_L/P_{cr}=1.5$. Figure 2(b) displays the initial beacon intensity profile for comparison. In Fig. 2(d) $|R|=0.96$, which, by visual inspection, reproduces the beacon profile more closely than Fig. 2(c) where $|R|=0.28$. However, as we will see below, the degree of reciprocity cannot be judged solely by similarity of the intensity profiles.

In Fig. (3) the quantities $(P_L/P_{cr})^{-2}\text{Re}\langle R(z_T)-1\rangle$ and $(P_L/P_{cr})^{-1}\text{Im}\langle R(z_T)\rangle$ are plotted as a function of $\sigma_r^2$ for three different powers. The curves nearly overlap, illustrating the $P_L/P_{cr}$ scaling. Both the real and imaginary components first drop; then, counterintuitively, increase with turbulence strength. A rough scaling can be derived to explain this behavior. The total nonlinear phase acquired by the beam can be approximated as $|k\int n_2 I_L(\mathbf{x})dz|\sim 2(P_L/P_{cr})(z/Z_R)\equiv 2\alpha$. If this phase is small, we can condense it into a single screen applied to the beam at the transmitter. Using this approximation and reciprocity, one can show

$$\langle R(z)\rangle-1\approx -\frac{4\alpha}{\pi w^2}\left[i\int\langle \hat{I}_R^2(\mathbf{r})\rangle d\mathbf{r}+\alpha\int\langle \hat{I}_R^3(\mathbf{r})\rangle d\mathbf{r}\right] \qquad (5)$$

where $\hat{I}_R=I_R/I_0$ and $I_0$ is the beacon's peak intensity. Equation (5) reproduces the $P_L/P_{cr}$ scaling observed in Figs. (2) and (3). Unfortunately, completing the integrals in Eq. (5) requires knowledge of the 4th and 6th order statistics [15]. To progress, we use rough dimensional arguments: $\int\langle \hat{I}_R^2(\mathbf{r})\rangle d\mathbf{r}\sim w_R^2\langle\hat{I}_R^2\rangle\sim w_R^2(1+\sigma_I^2)$ and $\int\langle \hat{I}_R^3(\mathbf{r})\rangle d\mathbf{r}\sim w_R^2\langle\hat{I}_R\rangle\langle\hat{I}_R^2\rangle\sim (1+\sigma_I^2)$, where $w_R^2$ is the average spot size at the receiver, $\sigma_I^2=\langle I_R^2\rangle/\langle I_R\rangle^2-1$ is the scintillation index, and we have used power conservation. This provides $\text{Im}\langle R\rangle\propto \alpha(1+\sigma_I^2)$ and $\text{Re}\langle R-1\rangle\propto -\alpha^2(1+\sigma_I^2)$.

The above scaling suggests that the dip and rise in $\langle R\rangle$ with $\sigma_r^2$ result from the same behavior observed in the scintillation index [14,15]. In the weak turbulence regime, the spatial phase distortions increase with turbulence strength. This, in turn, enhances the irradiance fluctuations causing $\sigma_I^2$ to grow. The initial drop in $\langle R\rangle$ can thus be interpreted as follows. At low powers, every beam 'ray' has a reciprocal beacon 'ray'. The rays travel through the turbulence along the same path, but in opposite directions. At high powers, the beam ray undergoes nonlinear refraction, continually deviating it from the path of its reciprocal counterpart. The random refraction experienced along the deviated path increases with turbulence strength, leading to greater, on average, path differences between the rays. This leads to a spatial phase difference and irradiance profile disjointedness at the target.

In the strong turbulence regime, the light becomes sufficiently spatially incoherent that the irradiance fluctuations saturate. The irradiance profile resembles that resulting from a collection of random sources [16]. The effective critical power for an incoherent beam is greater than that of a coherent beam, effectively causing the beam propagation to become more linear [11,17]. This linear-like propagation results in the increase of $\langle R \rangle$ with turbulence strength.

Departures of $|R|$ from unity can occur from both spatial phase differences and irradiance disjointedness between the beacon and beam. For many applications, such as power beaming and directed energy, the on-target quality of the irradiance profile, not the phase, is of primary interest. To examine this, we define a modified reciprocity metric

$$R_I(z) \equiv \frac{1}{2} \frac{\varepsilon_0 c}{[P_B(z)P_L(z)]^{1/2}} \int |E_B(\mathbf{r},z) E_L(\mathbf{r},z)| d\mathbf{r} \ . \qquad (6)$$

This metric accounts only for the amplitudes of the beacon and beam, and, as a result, satisfies the condition $R_I \geq |R|$. Figure (4) displays a comparison of $\langle |R| \rangle$ and $\langle R_I \rangle$ as a function of $P_L/P_{cr}$ for $\sigma_r^2 = 4.6$, the minimum of the reciprocity curve in Fig. (3). Figure (4) demonstrates that the loss in reciprocity is due primarily to phase differences between the beacon and beam and not irradiance disjointedness.

We have examined nonlinear reciprocity breakdown when AO phase correction is applied to high power laser beams propagating in random media. A metric, the overlap of a high power beam field and that of a beacon, was used to quantify reciprocity breaking. As an example, an ideal field-conjugation based AO implementation was applied to propagation through Kerr-nonlinear atmospheric turbulence. The reciprocity was found to drop with increasing beam power due primarily to spatial phase differences between the beacon and beam. This suggests AO correction can be effective in high power laser applications insensitive to phase quality.

**Funding** This work was supported by the Joint Technology Office and the Office of Naval Research.

**Acknowledgements** The authors would like to thank A. Ting, D. Gordon, and C.C. Davis for fruitful discussions.

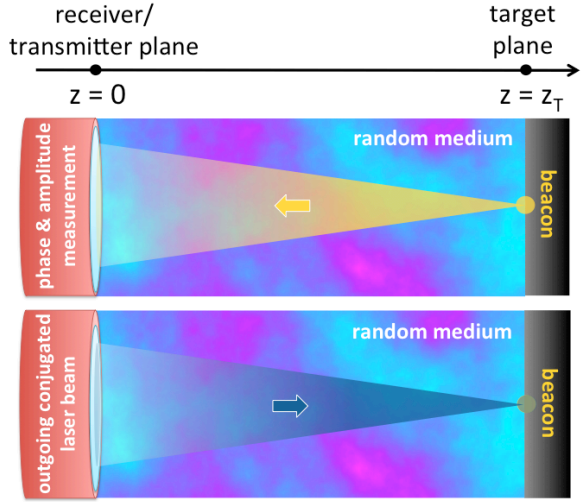

Figure 1. A beacon located on a target embedded in a random medium informs the phase and amplitude of a laser beam incident on the target.

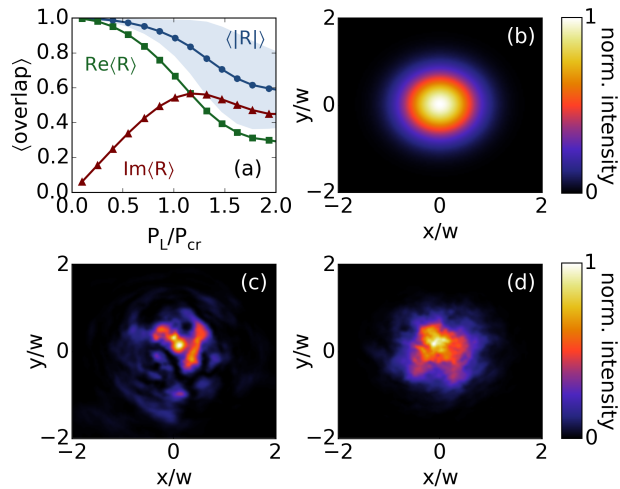

Figure 2. (a) Ensemble average of $R(z_T)$ as a function of $P_L/P_{cr}$ for $\sigma_r^2 = 6.8$. The dots, squares, and triangles show the means of $|R(z_T)|$, $\text{Re}[R(z_T)]$, and $\text{Im}[R(z_T)]$ respectively, and the swathes +/- the standard deviation of $|R(z_T)|$. (b) the initial beacon intensity profile on-target. (c) and (d) examples of low, $|R|=0.28$, and high, $|R|=0.96$, degrees of reciprocity, at $P_L = 1.5\ P_{cr}$. The reciprocity drops with increasing power due to nonlinear propagation of the beam.

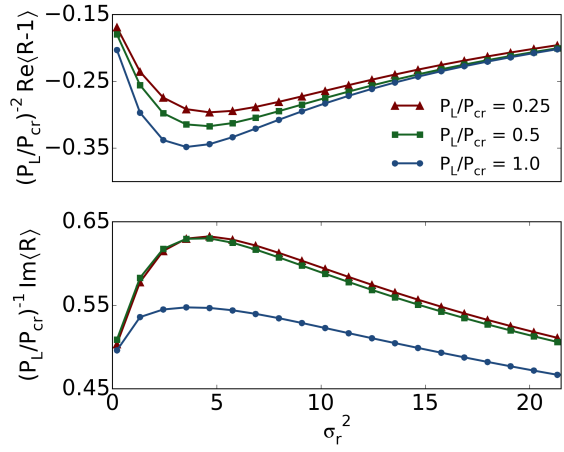

Figure 3. Ensemble average of $(P_L/P_{cr})^{-2} \text{Re}\langle R(z_T)-1\rangle$ and $(P_L/P_{cr})^{-1} \text{Im}\langle R(z_T)\rangle$ as a function of $\sigma_r^2$ for $P_L = 0.25\, P_{cr}$ red triangles, $P_L = 0.5\, P_{cr}$ green squares, and $P_L = 1.0\, P_{cr}$ blue circles.

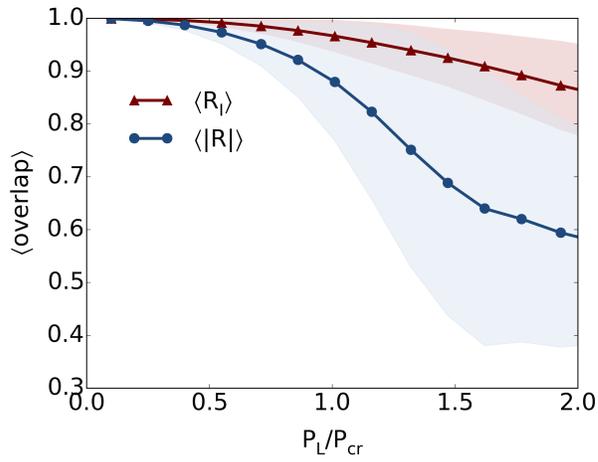

Figure 4 ensemble averages of $|R(z_T)|$, blue circles, and $|R_I(z_T)|$, red triangles, as a function of $P_L/P_{cr}$ for $\sigma_r^2 = 4.6$. The swathes indicate +/- the standard deviation.